\documentclass[useAMS,usenatbib,letters]{mn2e}

\usepackage[dvipdfmx]{graphicx}
\usepackage{amsmath}
\usepackage{journalnames}

\usepackage[T1]{fontenc}
\usepackage{aecompl}
\usepackage{yfonts}

\usepackage{color}

\title{The inefficiency of satellite accretion in forming extended star clusters}

\author[P. Bianchini et al.]{Paolo Bianchini$^{1,}$\thanks{E-mail:
bianchini@mpia.de}\thanks{Member of the International Max Planck Research School for Astronomy and Cosmic Physics at the University of Heidelberg, IMPRS-HD, Germany.}, Florent Renaud$^{2, 3}$, Mark Gieles$^{2}$, Anna Lisa Varri$^{4}$\\
$^{1}$Max-Planck Institute for Astronomy, Koenigstuhl 17, 69117 Heidelberg, Germany\\
$^{2}$Department of Physics, University of Surrey, Guildford GU2 7XH, UK\\
$^{3}$Laboratoire AIM Paris-Saclay, CEA/IRFU/SAp, Universit\'e Paris Diderot, F-91191 Gif-sur-Yvette Cedex, France\\
$^{4}$School of Mathematics and Maxwell Institute of Mathematical Sciences, University of Edinburgh, King's Buildings, Edinburgh EH9 3JZ, UK
}
 
\begin{document}

\date{Accepted 2014 November 4.  Received 2014 November 3; in original form 2014 October 15}

\pagerange{L1--L5} \pubyear{2014}

\maketitle

\label{firstpage}

\begin{abstract}
The distinction between globular clusters and dwarf galaxies has been progressively blurred by the recent discoveries of several extended star clusters, with size ($20-30$ pc) and luminosity ($-6<M_v<-2$) comparable to the one of faint dwarf spheroidals.
In order to explain their sparse structure, it has been suggested that they formed as star clusters in dwarf galaxy satellites that later accreted onto the Milky Way. If these clusters form in the centre of dwarf galaxies, 
they evolve in a tidally-compressive environment where the contribution of the tides to the virial balance can become significant, and lead to a super-virial state and subsequent expansion of the cluster, once removed.
Using $N$-body simulations, we show that a cluster formed in such an extreme environment undergoes a sizable expansion, during the drastic variation of the external tidal field due to the accretion process.
However, we show that the expansion due to the removal of the compressive tides is not enough to explain the observed extended structure, since the stellar systems resulting from this process are always more compact than the corresponding clusters that expand in isolation due to two-body relaxation. We conclude that an accreted origin of extended globular clusters is unlikely to explain their large spatial extent, and rather favor the hypothesis that such clusters are already extended at the stage of their formation.

\end{abstract}


\section{Introduction}

In recent years, a number of low-luminosity stellar systems was discovered around galaxies populating the transition region between low-luminosity dwarf spheroidal galaxies (dSphs) and globular clusters (GCs) in the luminosity-size parameter space. These stellar systems, with a luminosity between $-6<M_v<-2$ and a half-light radius of about $\approx20-30$ pc, are often referred to as extended clusters. In the Milky Way (MW) the known objects are Pal~4, Pal~14, AM~1 \citep{Mackey2005,HarrisCat2010}, and several extragalactic ones \citep{Brodie2002,Huxor2014,Mackey2013}. They are preferentially found in the outer halos of galaxies and are characterized by a more diffuse structure than typical GCs of similar luminosity. Therefore, their intermediate size, between the regime of dSphs galaxies and the bulk of GCs, makes them intriguing objects, crucial for the understanding of the differences and similarities of the formation of low-mass stellar systems.

Extended clusters have half-mass relaxation times slightly larger than their age, which makes it difficult to explain their sizes by expansion driven by two-body relaxation \citep{Gieles2011}.
The debate on the nature of these objects was recently enhanced by the discovery of the peculiar MW extended stellar system Laevens1/Crater (half-light radius between $20-30$ pc), whose classification as a GC or a dSph is still debated \citep{Laevens2014,Belokurov2014}.
Moreover, the complexity of their internal dynamics is not fully understood yet. \citet{Frank2014} found the surprising evidence of mass segregation in Pal~14. The current relaxation time of this extended cluster exceeds the Hubble time and therefore is too long to explain the settling of segregation only through dynamical evolution. The combined effect of primordial mass segregation and dynamical evolution could explain the structure of this cluster \citep{Haghi2014}. Therefore, further studies of the internal properties of such stellar systems are indeed crucial to unveil their origin.

The presumed origin of such extended clusters includes three main mechanisms: 1) they genuinely formed extended (this requires a formation environment with a high Mach number, which is appropriate to dwarf galaxies; \citealp{Elmegreen2008}); 2) they formed through the merging of two or more star clusters \citep{Fellhauer2002,Assmann2011}; 3) they were born as compact star clusters and later expanded due their peculiar environmental-driven evolution; this hypothesis will be tested in this study. Evolutionary processes that are often assumed to cause an expansion are strong tidal shocks \citep{Spitzer1958,Ostriker1972} or accretion process onto the MW halo \citep{Mackey2004,Miholics2014}.
The latter mechanism assumes that these clusters formed in dwarf-like satellite galaxies that later were stripped and accreted onto the halo of the host galaxy. We note that accretion of dwarf galaxies is a process commonly used to explain the structural properties of GC systems. In fact, there is growing evidence that a significant fraction of MW GCs are accreted systems, while the remaining formed in-situ in the early phase of galaxy formation \citep{Marin-Franch2009,Forbes2010,Leaman2013}. This is further supported by the spatial coincidence of outer halo GCs with stellar streams and overdensities \citep{Mackey2010}.

In this work, we focus on the possibility that extended clusters formed in the context of an accretion event, testing if  their observed extended sizes can be explained by the structural adjustment of the clusters to the time-dependent tidal field. We evaluate this hypothesis using $N$-body simulations considering the case of a cluster formed in the central regions of dwarf-like galaxies, where it experiences a compressive tidal field, which is then switched off to mimic the accretion of the cluster onto a MW-like galaxy. Compressive tides provide an extreme environment that enables the cluster to acquire an excess of kinetic energy with respect to its potential energy.
A complementary approach has been followed by \citet{Webb2014} and \citet{Miholics2014} for the study of the evolution of the size of clusters embedded in less extreme (non-tidally compressive) environments.

The paper is organized as follows: first we introduce the compressive tidal environment typical of cored regions of galaxies and show with analytical calculations that an accretion event causes an expansion of stellar systems. Then, in Section~3 we present our $N$-body simulations and we discuss our results. Finally, we report our conclusions in Section~4.

\section{Expansion of clusters due to satellite accretion}
\subsection{Compressive tides in galaxy cores}
\label{compressive_plummer}

Let's consider a gravitational potential $\phi$ embedding a star cluster. The associated tidal tensor can be written as the second space derivative of the potential \citep{Renaud2008}
\begin{equation}
T^{ij}=-\partial^i\partial^j\phi.
\end{equation}
The tides are fully compressive if all eigenvalues $\lambda_i$ of this tensor are negative, and extensive if at least one eigenvalue is positive. One typical environment of compressive tides is the central region of galaxies with cored density profiles. For dwarf galaxies, recent studies show that cored profiles are favored over cuspy ones \citep[e.g.][]{Walker2011}. In the case of a Plummer potential with characteristic radius $r_0$ fully compressive tides are found in the central region, delimited by $r<r_0/\sqrt{2}$  \citep{Renaud2010}.

\subsection{Proof of the principle}
We wish to describe the expansion of a stellar system due to a time-varying tidal field. We consider a globular cluster initially embedded in an isotropic compressive tidal field that is then instantaneously removed (impulsive approximation, \citealp{Spitzer1958}) to mimic the accretion event of the satellite that hosts the cluster in its centre. In this section, we follow the procedure outlined in \citet{Hills1980}.

Let's consider an isotropic tidal field such that the eigenvalues of the tidal tensor are  equal, $\lambda_i=\lambda,$ and negative (fully compressive tides). The initial energy of the cluster embedded in such field is \citep[][his eq. E.12]{Renaud2010}
\begin{equation}
E_0=\frac{1}{2}M_c\sigma^2-\frac{GM_c^2}{2r_v}-\frac{1}{2}\lambda\alpha M_c\,r_t^2
\end{equation}
where the last term describes the energy due to the compressive tides, $M_c$ the total mass of the cluster, $\sigma$ its velocity dispersion, $r_v$ the virial radius, $r_t$ the radius where the density of the cluster drops to zero, and $\alpha$ depends on the mass profile of the cluster and is defined as $\alpha=1/(M_c\,r_t^2)\int^{M_c}_0 r^2\, dm$.
We assume that the system is virialised before the compressive tides are removed \citep[][his eq. E.13]{Renaud2010}
\begin{equation}
M_c\sigma^2-\frac{GM_c^2}{2r_v}+\lambda\alpha M_c\,r_t^2=0.
\label{eq:virial}
\end{equation}
Since $\lambda$ is negative, we see that the velocity dispersion of the cluster is higher than what expected from a no-tide case.

In the impulsive approximation, we assume that both velocity dispersion and the radii of the cluster remain unchanged when the tidal field is instantaneously removed. This approximation is justified by the fact that the time scale in which the stripping of a dwarf-galaxy occurs is small compared to its internal dynamical time scale. The new energy of the system is therefore
\begin{equation}
E_1=\frac{1}{2}M_c\sigma^2-\frac{GM_c^2}{2r_v}
\label{eq:finalenergy}
\end{equation}
and, by using equation \ref{eq:virial}, we get
\begin{equation}
E_1=-\lambda\alpha M_cr_t^2-\frac{1}{2}M_c\sigma^2.
\end{equation}
The cluster remains bound ($E1 < 0$) for
\begin{equation}
\lambda>-\frac{1}{2}\frac{\sigma^2}{\alpha r_t^2}.
\label{eq:impulsive}
\end{equation}
Therefore, if the system is embedded in too strong compressive tides (very negative values of $\lambda$), its energy exceeds the reference level which would allow it to remain bound after the impulsive change. The cluster is then ``super-virialised" and is unbound when the tides are switched off. We here consider unbound when the total energy is positive, in reality the core can remain bound and stars in the outer part escape with velocities larger than the escape velocity.

After turning off the tidal field, the cluster settles in a new equilibrium state on a dynamical time scale, with a new radius $r_v'$ and a new velocity dispersion $\sigma'$. Neglecting any mass loss (i.e. constant $M_c$ over a dynamical time) the final state is described by the virial equation 
\begin{equation}
M_c\sigma'^2-\frac{GM_c^2}{2r'_v}=0,
\end{equation}
and a total energy
\begin{equation}
E=-\frac{GM_c^2}{4r'_v}.
\label{eq:newenergy}
\end{equation}
Using equation \ref{eq:virial}, \ref{eq:finalenergy} and \ref{eq:newenergy} we obtain a relation between the virial radius $r_v'$ at the final state and the initial radius $r_v$
\begin{equation}
r_v'=r_v\left(1+\frac{2\,\lambda\,\alpha\,r_t^2\,r_v}{G\,M_c}\right)^{-1}.
\label{eq:radiusimpulsive}
\end{equation}
In the case of compressive tides (i.e. $\lambda<0$), the final virial radius $r'_v$ is always larger than the initial $r_v$. The cluster therefore expands after the tidal field has been switched off.

In a realistic case we would expect that the expansion would affect more the stars in the outer part of the cluster. These stars will escape with nonzero velocities, taking away a large fraction of the energy gained during the compressive phase. This could indeed reduce the expansion. Therefore we discuss more detailed simulations in the next section. 


\section{$N$-body simulations}
\label{sec:main}

We now study the problem with $N$-body simulations, using Nbody6tt \citep{Renaud2011} based on Nbody6 \citep{Aarseth2003}. Nbody6tt gives the possibility to add to the regular forces the effect of an arbitrary time-dependent tidal field. We use the Graphics Processing Unit  (GPU) enabled version of \citet{Nitadori2012} and compute the simulations using the GPU cluster at the University of Surrey.

Our fiducial initial conditions for the cluster consist of 4096 particles drawn from a Plummer sphere. 
Since our investigation focuses specifically on the effects of an abrupt variation of the tidal environment of the star cluster, we consider exclusively equal-mass models, in the absence of stellar evolution. The compressive tides are given by the central region of another Plummer potential, mimicking the cored potential well in the center of a dwarf galaxy (see Sect. \ref{compressive_plummer}). The tidal tensor is computed analytically and fed to Nbody6tt. The tides are switched off to simulate the stripping of the dwarf galaxy as a consequence of the accretion process onto the MW. For simplicity, the potential of the MW is not considered: the clusters experience the tides from the dwarf galaxy alone, which are then fully removed.\footnote{
We note that our low-N cluster evolves to lower densities than real clusters. This is because in the expansion phase clusters evolve to a constant ratio of relaxation time over age, such that the clusters density is proportional to $N^2$. From Eq.~(9) we see that the expansion factor is smaller for cluster with higher density. This means that more massive clusters are less affected by the removal of the tides than the clusters in our $N$-body models.}

In the following sections we present the results of the long-term dynamical evolution of the cluster, considering different initial densities, tidal field strengths, transitions between compressive tides and isolation (impulsive or adiabatic transitions), and in orbit inside the dwarf galaxy potential. We consider as bound stars those with $E<0$ \citep{Renaud2011}, where the energy is given by the sum of the potential and kinetic energy, E=W+K.  
\begin{figure}
\centering
\includegraphics[width=0.45\textwidth]{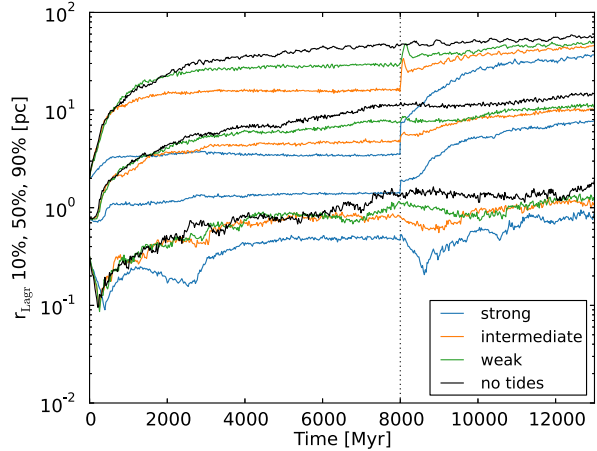}
\caption{Time evolution of the Lagrangian radii (enclosing 10\%, 50\% and 90\% of the total bound mass) of a cluster in three compressive tidal fields, 
labeled as weak (green lines), intermediate (orange lines) and strong (blue lines) tidal field. When the compressive tides are switched off at 8 Gyr (vertical line), the cluster expands. However, the expansion is not enough to generate objects more extended than the one evolved in isolation (black~lines).}
\label{fig:r_field_strength}
\end{figure}
\begin{figure}
\centering
\includegraphics[width=0.45\textwidth]{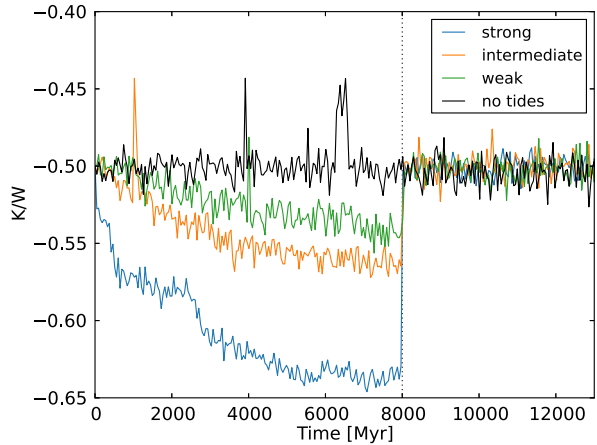}
\caption{Time evolution of the ratio of kinetic and potential energy $K/W$ of the runs presented in Fig. \ref{fig:r_field_strength}. The clusters are initialized with $K/W = -0.5$, i.e. neglecting the tides, but subsequently adjust quickly to it (see especially the case with strong tides).  When compressive tides are switched off, the cluster is in a ``super-virialized" state and tends to retrieve a virialized one by expanding. The peaks correspond to the formation of binaries.}
\label{fig:K_W_field_strength}
\end{figure}

\subsection{Evolution in the centre of a compressive tidal potential}
\label{no orbit}

\begin{table}
\begin{center}
\begin{tabular}{lllccc}
\hline\hline
Model &N& $r_v$ &$\mathcal{R}_{10\%}$&  $\mathcal{R}_{50\%}$ &  $\mathcal{R}_{90\%}$\\
\hline
W\_4k &4096 &1 pc&  0.53 & 0.75 & 0.80\\
I\_4k & 4096 &1 pc& 0.68 & 0.74 & 0.76\\ 
S\_4k & 4096 &1 pc&  0.40 & 0.46 & 0.56 \\
I\_8k & 8192 &1 pc& 1.01 & 0.90 & 0.83 \\
C\_4k & 4096 &0.4 pc& 1.01 & 0.65 &0.52\\
C\_4k\_1 & 4096&1 pc&  0.92& 0.77 &0.69\\
C\_4k\_2 & 4096 & 2.5 pc& 0.66 & 0.60 & 0.84\\
I\_4k\_ad &4096 &1 pc& 0.63 & 0.72 & 0.72\\

\hline
\end{tabular}
\caption{Ratios, $\mathcal{R}_{10,50,90}$, of the 10\%, 50\% and 90\% Lagrangian radii at 10 Gyr of a cluster evolved in compressive tides and then released to isolation at 8 Gry and the corresponding one evolved in isolation. The names of the simulations indicate the strength of the compressive tidal field (W, I, S, for weak, intermediate or strong; see Sect. \ref{no orbit}) or a circular orbit at 250 pc from the dwarf galaxy centre, indicated with C. The initial number of particles N and virial radius $r_\mathrm{v}$ are reported. For I\_4k\_ad the transition between the regime of compressive tides and isolation is adiabatic (see Sect. \ref{sec:imp_ad}). Clusters evolved in compressive tides are always less extended than the one evolved in isolation.}
\end{center}
\label{tab:runs}
\end{table}

The first case we explore consists in a cluster with virial radius $r_v=1$ pc.
We aim to study the dependence of the cluster's evolution on the strength of the tidal field. For this reason we place the cluster with no orbital motion in the centre of a dwarf galaxy with total mass $M=10^8M_\odot$ and scale radius $r_0=$ 1000~pc, 500~pc, 100~pc (typical values for dwarf galaxies masses and characteristic radii, \citealp{McConnachie2012}). The tidal field experienced is thus compressive, constant and isotropic. We label these three compressive tidal fields as weak, intermediate and strong. The compressive tides are switched off at 8~Gyr. At this time the clusters have already undergone core collapse and have expanded to their maximum extent (see Fig. \ref{fig:r_field_strength}).

In Fig. \ref{fig:r_field_strength} we show the time evolution of the 10\%, 50\% and 90\% Lagrangian radii of the cluster (i.e., the radii containing 10\%, 50\% and 90\% of the bound mass, respectively) for the strong, intermediate and weak compressive tidal fields. The evolution of the tidally perturbed model is compared to the one of an isolated cluster. While the compressive tides are present, the spatial extent of the cluster depends on the strength of the tidal field: the stronger the field, the more compact the cluster is.
The initial setup of the clusters ensures equipartition between kinetic and internal potential energies, and neglects the tidal term ($K/W = -0.5$). During the first phase, the negative energy brought by the compressive tides is balanced by an excess of kinetic energy with respect to the sole internal potential ($K/W < -0.5$). The time evolution of the ratio $K/W$ is depicted in Fig. \ref{fig:K_W_field_strength}.

When the compressive tidal field is switched off (8 Gyr), the cluster experiences an expansion, that is larger for stronger tides. We note that the expansion is significant for the 50\% and 90\% Lagrangian radii, while is negligible for the inner 10\% Lagrangian radius, confirming that tides only affect the outer regions of clusters.
We compare the spatial structure of the clusters evolved in compressive tides with the corresponding clusters evolved in isolation. From Fig. \ref{fig:r_field_strength}, it is clear that the expansion due to the abrupt variation of the tidal environment fails to produce objects that are more extended than the isolated cluster. We test this conclusion using a wide range of initial conditions including different initial cluster densities (initial virial radius of $r_v=0.4$, 1, 2.5 pc, initial number of particles $N=4096$, 8192) and circular orbits for the cluster inside the compressive tidal region. Nonetheless, all star clusters that underwent such a process are always less extended than the one evolved in isolation (a summary of the runs is reported in Tab. \ref{tab:runs}).

\subsection{Impulsive vs. adiabatic tidal change}
\label{sec:imp_ad}
\begin{figure}
\centering
\includegraphics[width=0.45\textwidth]{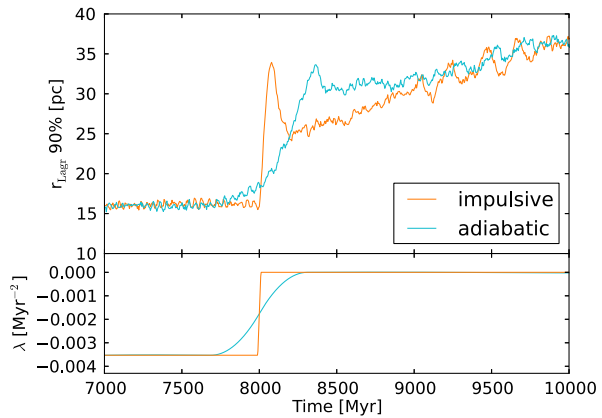}\,
\caption{Time evolution of the 90\% Lagrangian radius for a simulation with an impulsive transition between compressive tidal field and isolation (orange line) and one with an adiabatic transition lasting 600 Myr (cyan line). An intermediate strength of the compressive tidal field is used (see Fig. \ref{fig:r_field_strength}) and the corresponding eigenvalues $\lambda$ are shown in the lower sub-panel. The 90\% Lagrangian radii converge in short time to the same value.
}
\label{fig:trans_comp}
\end{figure}

So far we considered the case of an impulsive transition between the regime of compressive tidal field and no tidal field. We explore the effect of an adiabatic transition, occurring over a time-scale of 600 Myr.\footnote{This time-scale is the other extreme (compared to the impulsive transition), lasting longer than a typical accretion event.} Fig. \ref{fig:trans_comp} shows the evolution of the 90\% Lagrangian radius in the transition region only, for both an impulsive and adiabatic transition. The eigenvalues $\lambda$ associated to the time-dependent tidal tensors are shown in the lower sub-panel. Both simulations converge to the same radial extent in a few dynamical times, and no significant differences are present 1 Gyr after the transition (see also Tab. \ref{tab:runs}, model I\_4k and I\_4k\_ad).

\subsection{Observational surface density and velocity dispersion profiles}
\begin{figure}

\centering
\includegraphics[width=0.42\textwidth]{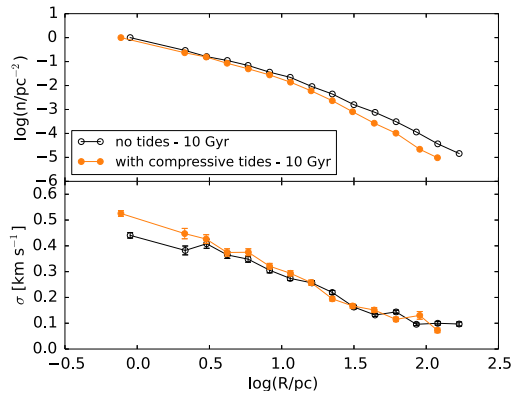}
\caption{Surface density profile (top panel) and line-of-sight velocity dispersion profile (bottom panel) at 10~Gyr of a cluster evolved in isolation (black line) and one evolved in intermediate-strength compressive tides (orange line) and released to isolation at $t=8$ Gyr. No significant structural differences are observable.
}
\label{fig:surf_profile}
\end{figure}

A further confirmation of the inefficiency of satellite accretion events in forming extended stellar systems, is given by the detailed analysis of the morphology and kinematics of our simulations. Fig. \ref{fig:surf_profile} displays the surface density profiles and the line-of-sight velocity dispersion profile as a function of the projected radius $R$ of snapshots at 10 Gyr of a cluster evolved completely in isolation and a cluster evolved initially in intermediate-strength compressive tides.\footnote{The profiles are constructed by stacking three snapshots around 10 Gyr, assuring to have a number of stars $>100$ per radial bin.} Despite the expansion imprinted by the time-dependent tidal field, the observational profiles do not show significant differences.\footnote{Note that the number of particles of the two clusters at 10 Gyr is comparable ($N_\mathrm{isolated}=1619$ and $N_\mathrm{tides}=1413$) and the measured half mass radii are $r_m=11.49$ pc and $r_m=8.16$ pc, for isolated and compressive tides case, respectively.}

\section{Conclusions}

We tested the possibility that extended clusters form originally in the dense central regions of dwarf galaxies and later expand due to a time-variation of the tidal field induced by the accretion of the host dwarf galaxy. In the core regions of dwarf galaxies, the clusters experience a regime of compressive tides that produces an excess of kinetic energy with respect to the internal potential energy. When the dwarf galaxy is accreted onto the MW, the clusters are released in the outer halo and expand due to this excess of energy.

We find that the expansion imprinted to the clusters does not give origin to objects that are more spatially extended than systems that have always evolved in isolation. We tested our conclusion exploring different initial densities for the clusters, orbits inside the core of the dwarf galaxy, strengths of the compressive tides and time variations of the transition from the regime of compressive tides to isolation.

We note here that our result can be considered as conservative. In fact, we tested the most extreme case in which the cluster originally resides in the central compressive region of a dwarf galaxy and is then released in isolation due to the accretion event. In a more realistic case, the accretion process of the dwarf galaxy would bring the cluster in the (extensive) tidal potential of the host galaxy. This would set a natural boundary for the spatial extent of the cluster (Lagrange surface), that would further limit its expansion. 
Moreover, by setting compressive tides, we push further the results of \citet{Miholics2014} who found that the (less extreme) case of tidally extensive tides does not lead to accreted clusters more extended than those in isolation.

We conclude that an accreted origin of outer halo extended clusters is unlikely to explain their large spatial extent. For this reason, these stellar systems could have genuinely formed extended or could have experienced an enhanced expansion due to some internal dynamical mechanism (e.g. the interplay of primordial mass segregation and dynamical relaxation, \citealp{Haghi2014}). Alternatively, observations of the radius of extended clusters could be biased by unbound stars that have already been stripped, and wonder in its vicinity along its orbit \citep{Kuepper2010}.

\section*{Acknowledgments}
This work was carried out as part of the ISIMA program on Gravitational Dynamics held in July 2014 at CITA, Toronto. We acknowledge Pascale Garaud for the organisation, the financial support and for the stimulating environment provided by all the participants. PB acknowledges support from Heidelberg Graduate School for Fundamental Physics. FR acknowledges support from the European Research Council (ERC) through grant ERC-StG-257720. MG acknowledges financial support from the Royal Society in the form of a University Research Fellowship (URF) and an equipment grant used for the GPU cluster in Surrey. MG and FR acknowledge support from the ERC (ERC-StG-335936, CLUSTERS) and ALV from RCE1851. Finally, all authors thank Sverre Aarseth and Keigo Nitadori for making the GPU version of NBODY6 available, and Dave Munro for the hardware support at the University of Surrey.

\bibliographystyle{mn2e} 
\bibliography{biblio} 

\begin{thebibliography}{31}
\expandafter\ifx\csname natexlab\endcsname\relax\def\natexlab#1{#1}\fi

\bibitem[{{Aarseth}(2003)}]{Aarseth2003}
{Aarseth} S.~J., 2003, {Gravitational N-Body Simulations}

\bibitem[{{Assmann} {et~al}\mbox{.}(2011){Assmann}, {Wilkinson}, {Fellhauer},
  \& {Smith}}]{Assmann2011}
{Assmann} P., {Wilkinson} M.~I., {Fellhauer} M., {Smith} R., 2011, \mnras, 413,
  2606

\bibitem[{{Belokurov} {et~al}\mbox{.}(2014){Belokurov}, {Irwin}, {Koposov},
  {Evans}, {Gonzalez-Solares}, {Metcalfe}, \& {Shanks}}]{Belokurov2014}
{Belokurov} V. {et~al.}, 2014, \mnras, 441, 2124

\bibitem[{{Brodie} \& {Larsen}(2002)}]{Brodie2002}
{Brodie} J.~P., {Larsen} S.~S., 2002, \aj, 124, 1410

\bibitem[{{Elmegreen}(2008)}]{Elmegreen2008}
{Elmegreen} B.~G., 2008, \apj, 672, 1006

\bibitem[{{Fellhauer} \& {Kroupa}(2002)}]{Fellhauer2002}
{Fellhauer} M., {Kroupa} P., 2002, \aj, 124, 2006

\bibitem[{{Forbes} \& {Bridges}(2010)}]{Forbes2010}
{Forbes} D.~A., {Bridges} T., 2010, \mnras, 404, 1203

\bibitem[{{Frank} {et~al}\mbox{.}(2014){Frank}, {Grebel}, \&
  {K{\"u}pper}}]{Frank2014}
{Frank} M.~J., {Grebel} E.~K., {K{\"u}pper} A.~H.~W., 2014, \mnras, 443, 815

\bibitem[{{Gieles} {et~al}\mbox{.}(2011){Gieles}, {Heggie}, \&
  {Zhao}}]{Gieles2011}
{Gieles} M., {Heggie} D.~C., {Zhao} H., 2011, \mnras, 413, 2509

\bibitem[{{Haghi} {et~al}\mbox{.}(2014){Haghi}, {Hoseini-Rad}, {Zonoozi}, \&
  {K{\"u}pper}}]{Haghi2014}
{Haghi} H., {Hoseini-Rad} S.~M., {Zonoozi} A.~H., {K{\"u}pper} A.~H.~W., 2014,
  \mnras, 444, 3699

\bibitem[{{Harris}(2010)}]{HarrisCat2010}
{Harris} W.~E., 2010, arXiv:1012.3224

\bibitem[{{Hills}(1980)}]{Hills1980}
{Hills} J.~G., 1980, \apj, 235, 986

\bibitem[{{Huxor} {et~al}\mbox{.}(2014){Huxor}, {Mackey}, {Ferguson}, {Irwin},
  {Martin}, {Tanvir}, {Veljanoski}, {McConnachie}, {Fishlock}, {Ibata}, \&
  {Lewis}}]{Huxor2014}
{Huxor} A.~P. {et~al.}, 2014, \mnras, 442, 2165

\bibitem[{{K{\"u}pper} {et~al}\mbox{.}(2010){K{\"u}pper}, {Kroupa},
  {Baumgardt}, \& {Heggie}}]{Kuepper2010}
{K{\"u}pper} A.~H.~W., {Kroupa} P., {Baumgardt} H., {Heggie} D.~C., 2010,
  \mnras, 407, 2241

\bibitem[{{Laevens} {et~al}\mbox{.}(2014){Laevens}, {Martin}, {Sesar},
  {Bernard}, {Rix}, {Slater}, {Bell}, {Ferguson}, {Schlafly}, {Burgett},
  {Chambers}, {Denneau}, {Draper}, {Kaiser}, {Kudritzki}, {Magnier},
  {Metcalfe}, {Morgan}, {Price}, {Sweeney}, {Tonry}, {Wainscoat}, \&
  {Waters}}]{Laevens2014}
{Laevens} B.~P.~M. {et~al.}, 2014, \apjl, 786, L3

\bibitem[{{Leaman} {et~al}\mbox{.}(2013){Leaman}, {VandenBerg}, \&
  {Mendel}}]{Leaman2013}
{Leaman} R., {VandenBerg} D.~A., {Mendel} J.~T., 2013, \mnras, 436, 122

\bibitem[{{Mackey} \& {Gilmore}(2004)}]{Mackey2004}
{Mackey} A.~D., {Gilmore} G.~F., 2004, \mnras, 355, 504

\bibitem[{{Mackey} {et~al}\mbox{.}(2010){Mackey}, {Huxor}, {Ferguson}, {Irwin},
  {Tanvir}, {McConnachie}, {Ibata}, {Chapman}, \& {Lewis}}]{Mackey2010}
{Mackey} A.~D. {et~al.}, 2010, \apjl, 717, L11

\bibitem[{{Mackey} {et~al}\mbox{.}(2013){Mackey}, {Huxor}, {Martin},
  {Ferguson}, {Dotter}, {McConnachie}, {Ibata}, {Irwin}, {Lewis}, {Sakari},
  {Tanvir}, \& {Venn}}]{Mackey2013}
{Mackey} A.~D. {et~al.}, 2013, \apjl, 770, L17

\bibitem[{{Mackey} \& {van den Bergh}(2005)}]{Mackey2005}
{Mackey} A.~D., {van den Bergh} S., 2005, \mnras, 360, 631

\bibitem[{{Mar{\'{\i}}n-Franch} {et~al}\mbox{.}(2009){Mar{\'{\i}}n-Franch},
  {Aparicio}, {Piotto}, {Rosenberg}, {Chaboyer}, {Sarajedini}, {Siegel},
  {Anderson}, {Bedin}, {Dotter}, {Hempel}, {King}, {Majewski}, {Milone},
  {Paust}, \& {Reid}}]{Marin-Franch2009}
{Mar{\'{\i}}n-Franch} A. {et~al.}, 2009, \apj, 694, 1498

\bibitem[{{McConnachie}(2012)}]{McConnachie2012}
{McConnachie} A.~W., 2012, \aj, 144, 4

\bibitem[{{Miholics} {et~al}\mbox{.}(2014){Miholics}, {Webb}, \&
  {Sills}}]{Miholics2014}
{Miholics} M., {Webb} J.~J., {Sills} A., 2014, \mnras, 445, 2872

\bibitem[{{Nitadori} \& {Aarseth}(2012)}]{Nitadori2012}
{Nitadori} K., {Aarseth} S.~J., 2012, \mnras, 424, 545

\bibitem[{{Ostriker} {et~al}\mbox{.}(1972){Ostriker}, {Spitzer}, \&
  {Chevalier}}]{Ostriker1972}
{Ostriker} J.~P., {Spitzer}, Jr. L., {Chevalier} R.~A., 1972, \apjl, 176, L51

\bibitem[{{Renaud}(2010)}]{Renaud2010}
{Renaud} F., 2010, PhD thesis

\bibitem[{{Renaud} {et~al}\mbox{.}(2008){Renaud}, {Boily}, {Fleck}, {Naab}, \&
  {Theis}}]{Renaud2008}
{Renaud} F., {Boily} C.~M., {Fleck} J.-J., {Naab} T., {Theis} C., 2008, \mnras,
  391, L98

\bibitem[{{Renaud} {et~al}\mbox{.}(2011){Renaud}, {Gieles}, \&
  {Boily}}]{Renaud2011}
{Renaud} F., {Gieles} M., {Boily} C.~M., 2011, \mnras, 418, 759

\bibitem[{{Spitzer}(1958)}]{Spitzer1958}
{Spitzer}, Jr. L., 1958, \apj, 127, 17

\bibitem[{{Walker} \& {Pe{\~n}arrubia}(2011)}]{Walker2011}
{Walker} M.~G., {Pe{\~n}arrubia} J., 2011, \apj, 742, 20

\bibitem[{{Webb} {et~al}\mbox{.}(2014){Webb}, {Leigh}, {Sills}, {Harris}, \&
  {Hurley}}]{Webb2014}
{Webb} J.~J., {Leigh} N., {Sills} A., {Harris} W.~E., {Hurley} J.~R., 2014,
  \mnras, 442, 1569

\end{thebibliography}

 \label{lastpage}

\end{document}